\documentclass[aps,prl,amsmath,twocolumn,amssymb,floatfix,showpacs,superscriptaddress,nofootinbib,longbibliography]{revtex4-1}
\pdfoutput=1

\usepackage{mathtools}
\usepackage{braket}
\usepackage[dvipsnames]{xcolor}
\usepackage{float}
\usepackage{subfigure}
\usepackage{dsfont}
\usepackage{tikz}
\usepackage[colorlinks=true,linktoc=page,linkcolor=magenta,citecolor=Violet,urlcolor=purple]{hyperref}

\usepackage{multirow}
\usepackage{comment} 
\usepackage{svg}
\usepackage{braket}
\usepackage[dvipsnames]{xcolor}
\usepackage{subfigure}
\usepackage{tikz}

\usepackage[normalem]{ulem}

\hypersetup{
    pdfnewwindow=true,      
    colorlinks=true,       
    linkcolor=red,          
    citecolor=blue,        
    filecolor=blue,      
    urlcolor=blue      
}

\newcommand{\ie}{{i.e., }}
\newcommand{\eg}{e.g.}

\newcommand{\non}{\nonumber}

\newcommand{\UP}{\uparrow}
\newcommand{\DN}{\downarrow}

\newcommand{\mb}{\mathbf}

\newcommand{\mc}{\mathcal}
\newcommand{\tbf}{\textbf}

\mathchardef\mhyphen="2D 
\newcommand{\ua}{{\uparrow }}
\newcommand{\da}{{\downarrow }}

\newcommand\bea{\begin{eqnarray}}
	\newcommand\eea{\end{eqnarray}}
\newcommand\beq{\begin{equation}}  
	\newcommand\eeq{\end{equation}}

\definecolor{lime}{HTML}{A6CE39}

\DeclareRobustCommand{\orcidicon}{\hspace{-1mm}
	\begin{tikzpicture}
	\draw[lime, fill=lime] (0,0) 
	circle [radius=0.16] 
	node[white] {{\fontfamily{qag}\selectfont \tiny \,ID}};
	\draw[white, fill=white] (-0.0525,0.095) 
	circle [radius=0.007];
	\end{tikzpicture}
	\hspace{-3mm}
}
\foreach \x in {A, ..., Z}{\expandafter\xdef\csname orcid\x\endcsname{\noexpand\href{https://orcid.org/\csname orcidauthor\x\endcsname}
			{\noexpand\orcidicon}}
}

\begin{document} 

\title{Emergent superconducting phases in unconventional $p$-wave magnets: Topological superconductivity, Bogoliubov Fermi surfaces and superconducting diode effect}

\author{Amartya Pal\orcidA{}}
\affiliation{Institute of Physics, Sachivalaya Marg, Bhubaneswar-751005, India}
\affiliation{Homi Bhabha National Institute, Training School Complex, Anushakti Nagar, Mumbai 400094, India}

\author{Paramita Dutta\orcidB{}}
\email{paramita@prl.res.in}
\affiliation{Theoretical Physics Division, Physical Research Laboratory, Navrangpura, Ahmedabad-380009, India}

\author{Arijit Saha\orcidC{}}
\email{arijit@iopb.res.in}
\affiliation{Institute of Physics, Sachivalaya Marg, Bhubaneswar-751005, India}
\affiliation{Homi Bhabha National Institute, Training School Complex, Anushakti Nagar, Mumbai 400094, India}


\begin{abstract}

The recent discovery of unconventional momentum-dependent magnetic orders has expanded the landscape of magnetism beyond conventional ferromagnetism and antiferromagnetism. Among them, $p$-wave magnets ($p$WMs) represent a novel class of odd-parity, non-collinear compensated magnetic order that generates spin-split electronic bands. In this work, our theoretical investigation establishes $p$WMs as a versatile platform for realizing intriguing superconducting phases including topological superconductivity (TSC), Bogoliubov Fermi surfaces (BFSs), and superconducting diode effect (SDE), within a unified microscopic framework. Employing a minimal model incorporating $p$-wave magnetic order, exchange coupling, and Zeeman fields, we perform a self-consistent mean-field analysis and uncover a rich phase diagram featuring unconventional finite-momentum Fulde–Ferrell (FF) and Larkin–Ovchinnikov (LO) superconducting phases. Remarkably, we also show that $p$WMs can undergo a transition to a TSC phase anchoring Majorana flat edge modes, a hallmark of two-dimensional TSCs, even without Rashba spin–orbit coupling and Zeeman field. Upon applying a Zeeman field, gapless FF and LO phases emerge with BFSs characterized by the appearance of finite zero-energy quasiparticle density of states. Furthermore, we demonstrate that SDE arises naturally in the asymmetric FF phase. Our analysis manifests that $p$WMs serve as a unique and novel platform to host TSC phase, gapless superconducting states, and non-reciprocal transport phenomena.
\end{abstract}


\maketitle



{\large{\tbf{Introduction }}}
\vskip -0.08cm
For decades magnetism has been broadly classified into two categories: ferromagnetism and antiferromagnetism, based on their spin ordering. Very recently, this dichotomy has been extended to include momentum-dependent magnetic orders with both even parity altermagnets~\cite{Smejkal_PRX_1,Smejkal_PRX_2,BhowalPRX2024,Mazin_PRBL_2023,Lee2024_PRL_MnTe,Biswas2026} along with odd parity $p$-wave magnets ($p$WMs)~\cite{Hellenes2024_pmagnets,Linder2025_pmagnet,Chakraborty2025,McNally2026}. These unconventional magnets exhibit spin-split electronic bands reminiscent of ferromagnets, while maintaining zero net magnetization, a property of antiferromagnets. The spin-splitting in these systems arises from the absence of combined $\mc{P}\mc{T}$-symmetry where $\mc{P}$ and $\mc{T}$ denote the parity and time-reversal symmetry (TRS) operators, respectively. Specifically, $p$WMs are characterized by $\mc{T}\mb{t}$-symmetry where $\mb{t}$ denotes the half lattice translation along with broken $\mc{P}$-symmetry~\cite{Linder2025_pmagnet,Chakraborty2025,Hellenes2024_pmagnets}. Notably, an important feature of the $p$WMs lies in their non-collinear magnetic order effectively mimics relativistic spin–orbit coupling (SOC) without relying on heavy elements~\cite{Hellenes2024_pmagnets,Linder2025_pmagnet,Chakraborty2025}. These properties have triggered growing interests and already been explored in some theoretical works~\cite{Sukhachov2025_pmagnet,Fukaya2025_Josephson_pwave,Ezawa2025b,Ezawa2025a,Nagae2025,Tim2025_SciPost}. Several materials such as CeNiAsO~\cite{Chakraborty2025}, ${\rm NiI_2}$~\cite{Song2025}, $\rm{Gd_3(Ru_{1-\delta}Rh_\delta)_4Al_{12}}$~\cite{Yamada2025} have been proposed as possible candidate materials, while their experimental realization still remains at an early stage~\cite{Song2025,Yamada2025}. 

The interplay between magnetism and superconductivity has long been a central theme in condensed matter physics giving rise to a variety of unconventional superconducting phases~\cite{Maple1990,Sigrist2009,Karki2013,Buzdin_2005, Bergeret2005}, especially, topological superconductivity~\cite{Kitaev_2001,Alicea_2012,Alicea2011_NatPhys,Beenakker2013search} and finite-momentum pairing states~\cite{LiangFu2021_PNAS,Kinnunen_2018} etc. While some of the phases have been explored in conventional magnets, and recently in altermagnets~\cite{Zhang2024,Chakraborty_2025_PRB,Chakraborty2025_PRL,Ghorashi2024PRL,Li2024,Li_PRBL_2023,Zhu2023,Zhang2017NatCommun,Mondal2025PRBL,Pal2025Flq_Josephson,Alam2025,Maeda2025,Chatterjee2025_AM,Ruthvik2025}, the role of $p$WMs in realizing such unconventional superconducting orders and their implications are yet to be understood.

Topological superconductivity (TSC) has attracted enormous attention due to its ability to host Majorana zero-energy modes, the charge-neutral quasiparticles obeying non-Abelian statistics and regarded as promising building blocks for fault-tolerant topological quantum computation~\cite{Kitaev_2001,Alicea_2012,Alicea2011_NatPhys,Beenakker2013search}. Existing platforms proposed for realizing Majorana modes, ranging from Rashba nanowires~\cite{LutchynPRL2010,Leijnse_2012,Das2012_NatPhys} and topological insulator surface states~\cite{Li2014} to altermagnets~\cite{Ghorashi2024PRL,Mondal2025PRBL,Pal2025Flq_Josephson,Alam2025}, all rely crucially on the SOC. Given that $p$WMs intrinsically emulate SOC, they provide a natural platform for studying TSC for which a microscopic picture is essential to identify the ground state.

Another hallmark of the interplay between magnetism and superconductivity is the finite-momentum pairing, in which Cooper pairs condense with a nonzero center-of-mass momentum and leads to periodically modulated superconducting order parameter in the real space~\cite{Kinnunen_2018}. Finite-momentum superconductivity is generally classified into the Fulde-Ferrell (FF) state~\cite{Fulde1964} and the Larkin-Ovchinnikov (LO) state~\cite{Larkin_1964}. In the FF phase, all Cooper pairs condense with identical center-of-mass momentum, while in the LO phase, there exists a time-reversal partner with center-of-mass momentum `$-\mb{q}$' for every Cooper pair with `$\mb{q}$'. Remarkably, the attention to the finite-momentum pairing has been renewed recently due to the possibility of the gapless superconducting phases hosting Bogoliubov Fermi surfaces (BFSs)~\cite{Zhu2021_BFS} and also for driving superconducting diode effects (SDEs)~\cite{LiangFu2022_PNAS,Daido2022_SDE_PRL}.

In literature, BFSs refer to an unconventional gapless superconducting phase in which the Bogoliubov quasiparticles coexist with Cooper pairs~\cite{Agterberg2017,Brydon2018,Setty2020PRB}. Unlike conventional Bardeen–Cooper–Schrieffer (BCS) superconductors exhibiting a full gap or nodal superconductors with vanishing quasiparticle density of states at zero energy~\cite{Sigrist2009}, systems with BFSs possess a finite elevated zero-energy density of states~\cite{Yuan2018,Setty2020NatComm,Lapp2020,Banerjee2022,Pal2024}. The dimension of BFSs is the same as the underlying normal state Fermi surface (FS) which is in sharp contrast to the BCS superconductors~\cite{Agterberg2017}. BFSs were theoretically proposed in spin-$3/2$ systems~\cite{Agterberg2017,Brydon2018,Timm2017,Oh2020,Oh2020,Timm2021a,Timm2021b,Dutta2021,Menke2019}, and later shown using BCS~\cite{Yuan2018,Banerjee2022,Cao2023,Wei2024,Serafim2024} and $d$-wave superconductors~\cite{Setty2020PRB,Setty2020NatComm,Pal2024,Pal2024_Thermal}, while they are experimentally observed in  Al/InAs heterostructures~\cite{Phan2022} and in ${\rm Bi_2Te_3}/{\rm NbSe_2}$ hybrid system driven by finite-momentum Cooper pairs~\cite{Zhu2021_BFS}. Given the potential to host finite-momentum Cooper pairs in $p$WMs and the possibility of coexistence of BFS and finite-momentum Cooper pairs, a natural question appears about the generation of BFSs in a gapless phase of $p$WMs. 

Then, in the context of the SDE, another phenomenon driven by finite-momentum Cooper pairs in broken inversion and time-reversal systems, the supercurrent exhibits a non-reciprocal behavior, \ie $|J_c^+|\ne |J_c^-|$ where, $J_c^+$ ($J_c^-$) denotes the critical supercurrent along the forward (reverse) direction~\cite{Nadeem2023,Jiang2022}. Unlike conventional semiconducting diodes which suffers from the Joule heating, SDE is attractive due to their dissipationless characteristics which are highly promising for device applications. SDE has been experimentally observed in Nb/Ta/V superlattices~\cite{Ando2020_SDE_Expt} and later found in other systems such as van der Waals heterostructure~\cite{Wu2022}, small-twist-angle trilayer graphene~\cite{Lin2022}, topological materials~\cite{Pal2022,Chatterjee2024}, and very recently, in altermagnets~\cite{Sourav2026}. It has also been theoretically studied in a wide range of systems including Rashba superconductors~\cite{Bergeret,Legg2022_SDE_MCA,Hasan2024,Bhowmik2025b}, magnetic ad-atoms~\cite{Bhowmik2025b,Samanta2025}, Josephson junctions~\cite{LiangFu2022_JDE,Legg2023,Debnath2024_diode,Debnath_2025,Debnath_2025AMDiode}, and also in altermagnets~\cite{Banerjee2024_PRB,Chakraborty2025_PRL,Pal2025_AM_diode,Constantin2026}. While the diode effect in $p$WM-based Josephson junction is studied very recently~\cite{Sharma2026_JDE_pmagnet}, microscopic analysis of SDE in bulk $p$WMs remains unexplored. 
Given that SDE is known to emerge only in the FF phase in the literature, while it is absent in the LO ~phase~\cite{LiangFu2022_PNAS,Daido2022_SDE_PRL,Banerjee2024_PRL}, exploring SDE in $p$WMs presents a natural and compelling direction.

\begin{figure*}
	\includegraphics[scale=0.66]{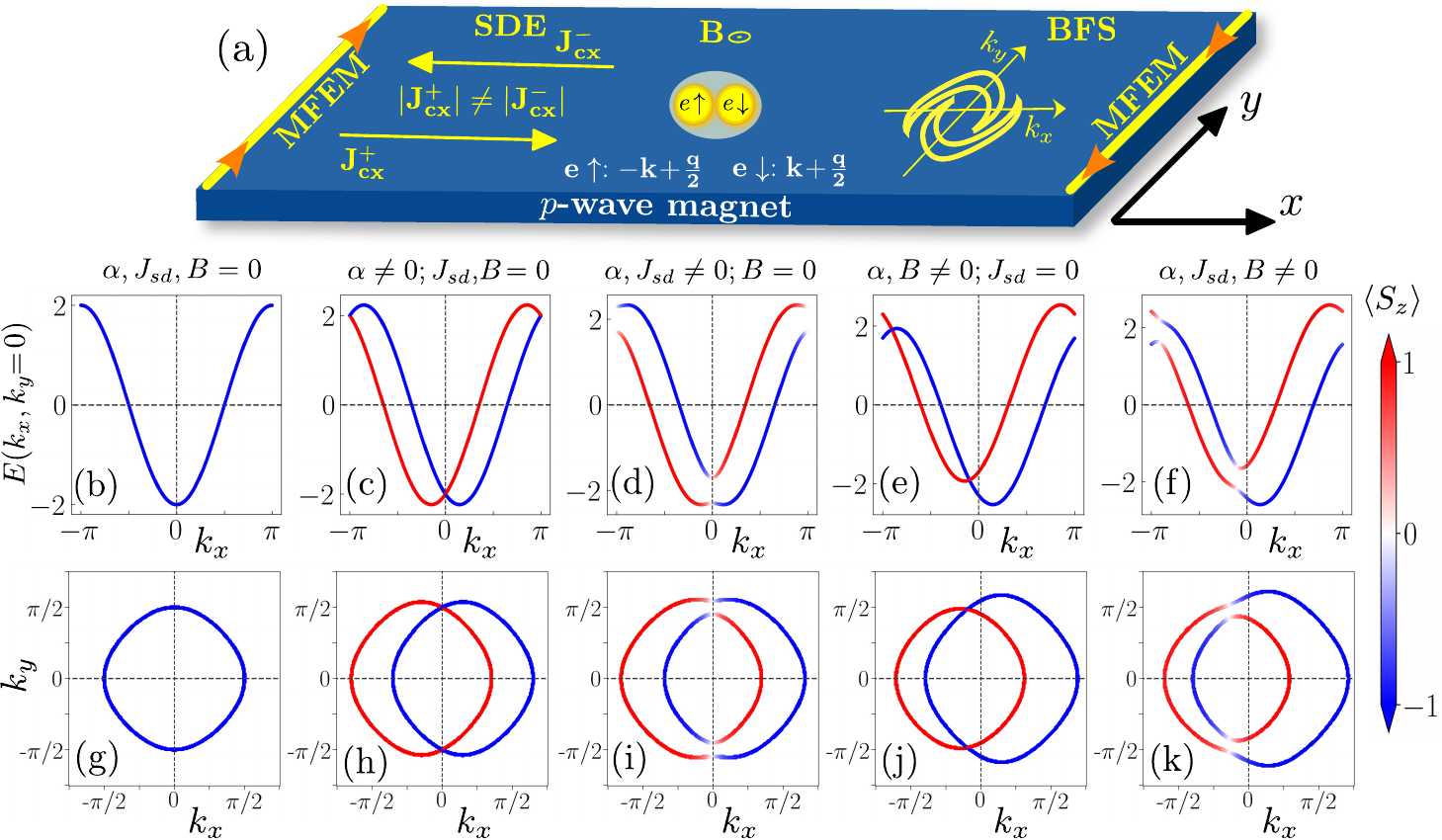} 
	\caption{\tbf{Illustration of the unconventional superconducting phases and the normal state spectral properties.} \tbf{(a)} Schematic illustration for possible phases in $p$WM in the presence of an external Zeeman field $B$, along with the MFEMs at the boundary. In panels, \tbf{(b)-(f) [(g)-(k)]} bulk energy-spectrum [Fermi surfaces] along with spin-polarization $\braket{S_z}$ (in units of $\hbar/2$) have been demonstrated. Other model parameter values 
	are chosen as \tbf{(b),(g)} $(\alpha,J_{sd},B)=(0,0,0)$, \tbf{(c),(h)} $(\alpha,J_{sd},B)=(t,0,0)$, \tbf{(d),(i)} $(\alpha,J_{sd},B)=(t,0.3t,0)$, \tbf{(e),(j)}$(\alpha,J_{sd},B)=(t,0,0.3t)$, \tbf{(f),(k)} $(\alpha,J_{sd},B)=(t,0.3t,0.3t)$ with $t=1$ and $\mu=-2t$.}
	\label{Fig_1}
\end{figure*}

With this wide range of motivation, we address the following intriguing questions: (i) is it possible to find TSC in $p$WMs without relying on Rashba SOC or any external Zeeman field? (ii) Is there any ground state for the FF or LO phase in $p$WMs hosting finite-momentum Cooper pairs? (iii) Can finite-momentum pairing induce BFSs in $p$WMs? (iv) Is it possible to generate SDE and how is it affected by the presence of BFSs? To answer these questions, we consider a $p$WM characterized by the odd-parity magnetic order and $sd$-exchange interaction~\cite{Linder2025_pmagnet} both in the absence and presence of Zeeman field. First, we analyse the spectral properties of the normal state Hamiltonian in detail to elucidate the role of the exchange coupling and Zeeman splitting. Introducing onsite attractive Hubbard interactions~\cite{Bhowmik2025a,Bhowmik2025b}, we then perform a fully self-consistent mean-field analysis allowing BCS, FF, and LO pairing channels. We demonstrate that the interplay between the exchange interaction and Zeeman field leads to a rich superconducting phase diagram, as shown in the diagrammatic representation of Fig.\,\ref{Fig_1}(a), hosting TSC with Majorana flat edge modes (MFEMs), gapped and gapless FF phases, and a gapless LO phase. We further illustrate that the finite-momentum pairing state drives the system into a gapless phase anchoring BFSs. Finally, we demonstrate the appearance of a tunable SDE in the FF phase, persisting even in the presence of BFSs. Our results establish $p$WM as a versatile platform for achieving unconventional superconducting phases with possible device application.
 
\vskip 0.1cm
{\large{\tbf{Results}}}

\tbf{Model.} We consider a two-dimensional (2D) tight binding Hamiltonian in the  momentum-space, $\mc{H}_{p{\rm WM}}$, that describes a $p$WM in the normal state with required symmetries, and apply an external Zeeman field $B$~\cite{Linder2025_pmagnet}. The total Hamiltonian takes the form:
\begin{equation}
	H_{\rm N} =\!\! \sum_{\mb{k}} \psi_{\mb{k}}^\dagger \mc{H}_{\rm N}(\mb{k}) \psi_{\mb{k}}=\!\!\sum_{\mb{k}} \psi_{\mb{k}}^\dagger [\mc{H}_{p{\rm WM}}(\mb{k}) \!+\! B\sigma_z] \psi_{\mb{k}}\ , \label{Eq.H_Nk}
\end{equation} 
where, 
\begin{eqnarray}
\mc{H}_{p{\rm WM}} (\mb{k}) &=& [-2t(\cos k_x + \cos k_y) -\mu]\sigma_{0}\tau_{0} + \alpha \sin k_x \sigma_z\tau_{0} \nonumber \\
&+& J_{sd} \sigma_x \tau_z\ , 
\end{eqnarray}
and $\psi_{\mb{k}} = \{c_{\mb{k}A\UP},c_{\mb{k}B\UP},c_{\mb{k}A\DN},c_{\mb{k}B\DN}\}^T$ with $c_{\mb{k}\gamma,s}$ ($c^\dagger_{\mb{k}\gamma,s}$) representing the annihilation (creation) operator for an electron with momentum $\mb{k}=\{k_x,k_y\}$ and spin $s = \{\UP, \DN\}$ in the orbital $\gamma = \{A, B\}$. The Pauli matrices $\sigma$ and $\tau$ act on the spin and orbital degrees of freedom, respectively. The model parameters $t$ and $\mu$ represent the nearest-neighbour hopping amplitudes and the chemical potential, respectively. Importantly, $\alpha$ and $J_{sd}$ parametrize the $p$-wave magnetic exchange orders which originates from the exchange interaction between the itinerant electrons and the localized non-collinear magnetic moments~\cite{Linder2025_pmagnet,Sukhachov2025_pmagnet}. Specifically, $\alpha$ corresponds to the spin-dependent hopping amplitude and $J_{sd}$ denotes the isotropic $sd$-coupling between itinerant electrons and localized moments, featuring partially spin-polarized bands. As discussed previously, $\mc{H}_{p{\rm WM}}(\mb{k})$ preserves the TRS i.e., $\mc{T}\mc{H}_{p{\rm WM}}(-\mb{k})\mc{T}^{-1} = \mc{H}_{p{\rm WM}}(\mb{k})$ where $\mc{T}=i\sigma_y \tau_x \mc{K}$ with $\mc{K}$ being the complex conjugation operator. Needless to say, the TRS is broken when the magnetic field $B$ is switched on. Here, the external magnetic field only couples with the spin degrees of freedom neglecting the orbital effects~\cite{Chakraborty2025_PRL}. We consider $\mu=-2t$ unless otherwise specified and set $t=1$ (in units of energy).

Focussing on the spectral features of the normal state to understand the role of each model parameters and the magnetic field, we first compute the bulk spectrum and the FS by diagonalizing $\mc{H}_{N}(k)$. Then, we compute the expectation value of the spin polarization along the $z$-direction i.e., $\braket{S_z}_{\mb{k}} = \frac{\hbar}{2} \braket{\sigma_z}_{\mb{k}} = \frac{\hbar}{2} \bra{u_{\mb{k}}}\sigma_z \ket{u_{\mb{k}}}$, where $\ket{u_{\mb{k}}}$ is the eigenvector of the Hamiltonian $\mc{H}_N(\mb{k})$ with eigenvalue $E(\mb{k})$. We depict the bulk spectrum $E(k_x,k_y=0)$ as a function of $k_x$ in Figs.\,\ref{Fig_1}(b)-(f) and the FS in the $k_x-k_y$ plane in Figs.\,\ref{Fig_1}(g)-(k) for various choices of model parameters. We also display the spin-polarization, $\braket{S_z}_{\mb{k}}$ (in units of $\hbar/2$) on top of bulk spectrum and FS in each plot. 

In Figs.\,\ref{Fig_1}(b) and (g), when $\alpha,J_{sd},B=0$, we observe the bulk bands to be four-fold degenerate along with the degenerate FS. 
When $\alpha\ne0$ (keeping $J_{sd},B=0$), the bulk bands as well as the FS split along $k_x$-direction lifting the four-fold degeneracy except at $k_x=0,\pi$ (see Figs.\,\ref{Fig_1}(c) and (h)). Note that, the FS remains spin-polarized as $\mc{H}_N(\mb{k})$ commutes with $\sigma_z$, i.e., $[\mc{H}_N(\mb{k}),\sigma_z]=0$. Now, if we consider both $\alpha,J_{sd} \ne 0$ (still $B=0$), the four-fold degeneracy at $k_x=0,\pi$ for the band as well as FS is lifted as shown in Figs.\,\ref{Fig_1}(d) and (i). The FS becomes partially spin-polarized in this parameter space since $[\mc{H}_N(\mb{k}),\sigma_z]\ne 0$ (see Fig.\,\ref{Fig_1}(i)). However, the TRS is still respected in the system which is reflected in the presence of Kramer's pairs maintaining $E(\mb{k},s)=E(-\mb{k},-s)$ where the spin $s=\braket{S_z}_{\mb{k}}$. 

So far, the magnetic field is switched off in our consideration. In the presence of a magnetic field, when $\alpha \ne0$ and $J_{sd}=0$, the system breaks TRS, lifting the Kramer's degeneracy i.e., $E(\mb{k},s)\ne E(-\mb{k},-s)$ as visible in the energy spectrum of Fig.~\ref{Fig_1}(e), whereas the corresponding FS becomes fully spin-polarized (see Fig.~\ref{Fig_1}(j)) as $[\mc{H}_N(\mb{k}),\sigma_z]=0$. Finally, when we turn on all the three components i.e., $\alpha,J_{sd},B\ne0$, the bulk bands as well as the FS (see Figs.\,\ref{Fig_1}(f) and (k)) are partially spin-polarized. Notably, the bands and FS become asymmetric with respect to $k_x$ indicating the possibility of finite-momentum pairing both in the absence and presence of $J_{sd}$ as long as $\alpha,B\neq 0$.  For any state with momentum and spin $\{\mb{k},s\}$, the partner state with $\{-\mb{k},-s\}$ is not available, restricting the formation of conventional BCS type pairing, which we discuss in the upcoming sections in more detail.

\tbf{Mean-field analysis.}
In order to investigate the possible superconducting orders in the normal state Hamiltonian of $p$WMs, we consider an attractive onsite electron-electron interaction of the form,
\begin{equation}
	H_U = -U\!\!\!\!\!\!\! \sum_{\mb{r},\gamma=\{A,B\}}\!\!\!\!\!\! n_{\mb{r}\gamma\ua}n_{\mb{r}_\gamma \da} \ ,
	\label{Eq. H_U_real_space}\,                                                
\end{equation}
where, $n_{\mb{r}\gamma s}=c^\dagger_{\mb{r}\gamma,s} c_{\mb{r}\gamma s}$ with $c^\dagger_{\mb{r}\gamma,s}$ being the electron creation operator at position $\mb{r}$ in the orbital $\gamma$ with spin $s$. Here, $U(>0)$ is the strength of attractive interaction. Origin of such electron-electron correlation can be intrinsic to the system~\cite{Sigrist2009,Kotegawa2014} or can be induced via the proximity effect~\cite{Beenakker1997,Buzdin_2005}. 
We can rewrite the interaction term in momentum space as, 
\begin{equation}
	H_U = -\frac{U}{N} \!\!\sum_{\mb{k_1},\mb{k_2},\mb{k_3},\gamma}\!\!\!\! c^\dagger_{\mb{k_1}+\mb{k_3} \gamma \ua} c^\dagger_{\mb{k_2}-\mb{k_3} \gamma \da} c_{\mb{k_2} \gamma \da} c_{\mb{k_1} \gamma \ua}\ ,
 \label{Eq. H_U_mtm_space}
\end{equation}
where, $N$ is the grid size of the Brillouin zone. Then, we perform a mean-field decomposition into the conventional BCS and also into the unconventional finite-momentum FF and LO pairing channels which is of our primary interest. For the finite-momentum pairing channel, we consider $\mb{k_2}=-\mb{k_1}+\mb{q}$ with $\mb{q}= \{ q_x,q_y\}$ being the center-of-mass momentum of the Cooper pair. The order parameters corresponding to these three pairing channels are defined as~\cite{Kinnunen_2018}: 
 \begin{eqnarray}
 	\Delta^{\rm BCS} &=& -\frac{U}{N} \sum_\mb{k} \braket{c_{-\mb{k} \gamma \da} c_{\mb{k}\gamma\ua}}\ , \\ 
 	\Delta^{\rm FF}_{\mb{q}} &=& -\frac{U}{N} \sum_\mb{k} \braket{c_{-\mb{k}+\frac{\mb{q}}{2} \gamma \da} c_{\mb{k} +\frac{\mb{q}}{2} \gamma\ua}}\ , \\ 
 	\Delta^{\rm LO}_{\mb{q}} &=& -\frac{U}{2N} \sum_\mb{k} [\braket{c_{-\mb{k}+\frac{\mb{q}}{2} \gamma \da} c_{\mb{k} +\frac{\mb{q}}{2} \gamma\ua}} + \braket{c_{-\mb{k}-\frac{\mb{q}}{2} \gamma \da} c_{\mb{k} -\frac{\mb{q}}{2} \gamma\ua}}]\ . \non \\
 \end{eqnarray}

Interestingly, the order parameter for the FF channel $\Delta^{\rm FF}_{\mb{q}}$ appears to be the most general one. It takes into account both the BCS and LO pairing channels, explicitly, $\Delta^{\rm BCS}=\Delta^{\rm FF}_{\mb{q=0}}$ and $\Delta^{\rm LO}_{\mb{q}}= (\Delta^{\rm FF}_{\mb{q}} + \Delta^{\rm FF}_{\mb{-q}})/2$. Thus, we only consider $\Delta^{\rm FF}_{\mb{q}}$ throughout the rest of the discussions to characterize the superconducting ground state. Also, throughout the rest of the manuscript, we drop the label $\gamma$ assuming equal magnitude of intra-orbital pairing amplitude.

The Bogoliubov-de Gennes (BdG) Hamiltonian after the mean-field decomposition becomes
\begin{equation}
	H_{\rm BdG}\!=\! \frac{1}{2} \sum_\mb{k} \Psi^\dagger_{\mb{kq}} \mc{H}_{\rm BdG} (\mb{k},\mb{q}) \Psi_
	{\mb{kq}} + \frac{2N}{U}| \Delta^{\rm FF}_{\mb{q}}|^2 \,+\,\, {\rm const.}
\end{equation}
where, $\Psi_\mb{kq} = (c_{\mb{k}+\frac{\mb{q}}{2}A\ua},c_{\mb{k}+\frac{\mb{q}}{2}B\ua}, c_{\mb{k}+\frac{\mb{q}}{2}A\da}, c_{\mb{k}+\frac{\mb{q}}{2}B\da},  c^\dagger_{-\mb{k}+\frac{\mb{q}}{2}A\ua},\\ c^\dagger_{-\mb{k}+\frac{\mb{q}}{2}B\ua},  c^\dagger_{-\mb{k}+\frac{\mb{q}}{2}A\da},c^\dagger_{-\mb{k}+\frac{\mb{q}}{2}B\da})^T$ is the Nambu spinor and 
\begin{equation}
	\mc{H}_{\rm BdG}(\mb{k},\mb{q}) = \begin{bmatrix}  
		\mc{H}_{\rm N}(\mb{k}+\frac{\mb{q}}{2}) & -i\sigma_y \Delta^{\rm FF}_{\mb{q}} \\  
		i\sigma_y \Delta^{\rm FF}_{\mb{q}} & -\mc{H}_{\rm N}^T(-\mb{k}+\frac{\mb{q}}{2}) 
	\end{bmatrix}		\label{Eq.H_BdG}
\end{equation}
with $\mc{H}_{\rm N}(\mb{k}+\frac{\mb{q}}{2})$ is the normal state Hamiltonian (see Eq.\,\eqref{Eq.H_Nk}).

\begin{figure*}
	\centering
	\includegraphics[scale=0.82]{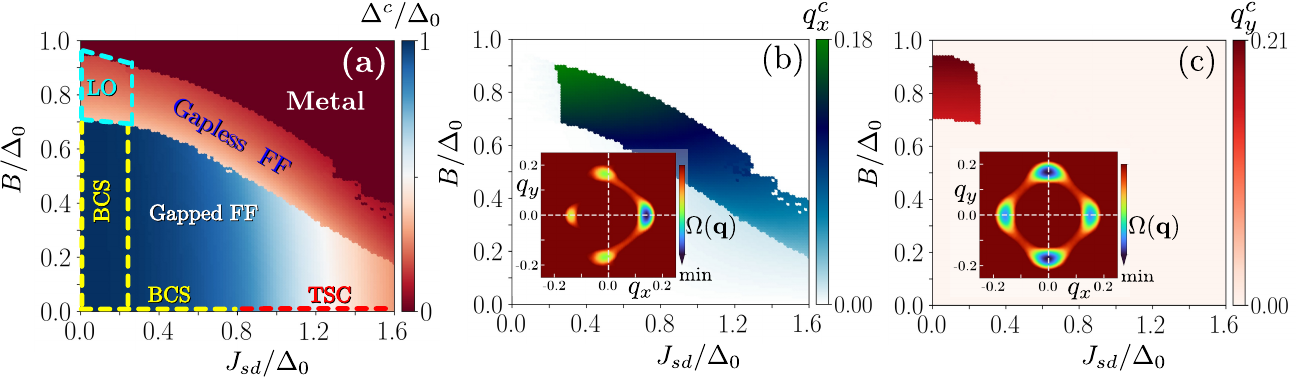} 
	\caption{\tbf{Phase diagram for the superconducting ground state:} Behavior of the \tbf{(a)} superconducting order parameter $\Delta^{\!c}/\Delta_0$, \tbf{(b)} $q_x^c$ and \tbf{(c)}  $q_y^c$  are depicted across the parameter space spanned by $J_{sd}$ and $B$. In panel \tbf{(a)}, we highlight the unconventional superconducting phases, TSC along with gapped FF, gapless FF, and LO pairing states. Inset of \tbf{(b)} and \tbf{(c)} display the condensation energy density, $\Omega(\mb{q},\Delta_{\mb{q}}^{\rm FF})$, in the $q_x\mhyphen q_y$ plane, corresponding to the gapless FF [$(J_{sd},B)=(0.5,0.7)\Delta_0$] and gapless LO  [$(J_{sd},B)=(0,0.75)\Delta_0$] state, respectively. }  	
	\label{Fig_2}
\end{figure*}

We now define the condensation energy density, $\Omega(\mb{q},\Delta_{\mb{q}}^{\rm FF})$, at the superconducting ground state as~\cite{Kinnunen_2018}, $\Omega(\mb{q},\Delta_{\mb{q}}^{\rm FF})= \mc{F}(\mb{q},\Delta_{\mb{q}}^{\rm FF}) - \mc{F}(\mb{q},0)$, where $\mc{F}(\mb{q},\Delta_{\mb{q}}^{\rm FF})$ and $\mc{F}(\mb{q},0)$ represent the free energy density at the superconducting and normal state of the system, respectively. At near zero temperature, the free energy density is computed using the relation~\cite{Coleman_2015}: $\displaystyle{\mc{F}(\mb{q},\Delta_{\mb{q}}^{\rm FF}) = \frac{1}{N} \sum_{\mb{k}, \mc{E}(\mb{k,q})<0}\!\!\!\!  \mc{E}(\mb{k,q}) + \frac{2N}{U}|\Delta^{\rm FF}_{\mb{q}}|^2}$ where $\mc{E}(\mb{k,q})$ is the energy eigenvalue of the BdG Hamiltonian (Eq.~\,\eqref{Eq.H_BdG}). The critical values of $\Delta^{\rm FF}_{\mb{q}}$, $q_x$, and $q_y$ denoted by $(\Delta^c,q_x^c,q_y^c)$ that minimize the condensation energy are computed by self-consistently capturing the energetically most favored state. 
In our numerical calculation, we set $U/2t=1.5$, $\mu/2t=-1$, $\alpha/t=1$ (unless otherwise specified) and consider a $500\times 500$ ($N=25\times 10^4$) momentum-space grid size. Changing these numerical 
values do not affect our main results qualitatively, unless the Hubbard strength is large for which the mean-field treatment will no longer remain valid. 

By investigating the superconducting ground state throughout the parameter space, we obtain a rich phase diagram demonstrated in Fig.~\ref{Fig_2} and provide a comprehensive analysis of all the phases. For $J_{sd}=B=0$, the superconducting ground state is given by $(\Delta^c,q_x^c,q_y^c)=(0.43t,0,0)$ which corresponds to the conventional BCS ground state hosting zero-momentum Cooper pairs ($s$-wave). We 
define $\Delta^c|_{J_{sd},B=0} \equiv \Delta_0$ and scale $\Delta^c$ by $\Delta_0$. We then 
analyse self-consistently computed $\Delta^c/\Delta_0, q_x^c$, and $q_y^c$ 
in the plane of $J_{sd}$ and $B$ as shown in Figs.\,\ref{Fig_2}(a), (b), and (c), respectively. We  identify various superconducting phases in Fig.\,\ref{Fig_2}(a) that includes conventional BCS, gapped FF, gapless FF, and gapless LO phases. 

When $J_{sd}=0$, then if we only vary the magnetic field $B$, then the system evolves from a conventional $s$-wave to a state with finite-momentum pairing (see Fig.\,\ref{Fig_2}(a)) but only along $y$-direction i.e., $q_x^c=0$ and $q_y^c\ne0$, as confirmed from Figs.\,\ref{Fig_2}(b) and (c). The variation of $\Omega(\mb{q})$ in this phase (see inset of Fig.\,\ref{Fig_2}(c)) indicates the existence of two degenerate ground states at $(q_x,q_y)=(0,\pm q_y^c)$, thus confirming it to be the finite-momentum pairing state in the LO channel. Interestingly, the transition from the BCS to the LO state takes place at a critical value of the Zeeman field $B^c=0.7\Delta_0\simeq \Delta_0/\sqrt{2}$ which corresponds to the Chandrasekhar-Clogston limit or Pauli limit~\cite{Closton1962,Chandrasekhar1962}. This limit refers to the critical strength of the external Zeeman field above which the conventional BCS state transforms to a normal metallic state. However, a finite-momentum pairing state can be stable even above this limit~\cite{Kinnunen_2018}, as reflected in the phase diagram. 

Then, when $J_{sd}\ne0$, we find the LO state is stable upto a certain finite value of $J_{sd}$ above which (for any finite strength of $B$) the FF pairing state becomes energetically more favorable with a finite Cooper pair momentum along $x$-direction ($q_x^c\ne0$) while $q_y^c=0$ (see inset of Fig.\,\ref{Fig_2}(b)).
Since, there is only one ground state, we confirm this to be the FF state.
As we increase the strength of $B$, the system remains in the FF pairing state however the spectral gap of the BdG Hamiltonian vanishes. This transition from the gapped to gapless FF pairing state is associated with a discontinuous change in $(\Delta^c,q_x^c)$  (see supplementary materials (SM) 
for details). Thus, we successfully uncover the possible finite-momentum pairing states in $p$WMs in the presence of an external Zeeman field. 

With the understanding of the finite-momentum pairing states, we next discuss the implications correponding
to these states. 
Interestingly, the system undergoes a transition to the TSC phase at $J_{sd}^c=0.8\Delta_0$ in the absence of the external Zeeman field as illustrated in Fig.\,\ref{Fig_2}(a) which we discuss in detail as follows.

\vskip 0.1cm
\tbf{Topological superconductivity:}

In order to establish the emergence of topological phase in our system, we systematically analyse the eigenvalues and eigenvectors of $\mc{H}_{\rm BdG}(\mb{k,q})$ within the bulk as well as at the boundary. 
When $B=0$, we can rewrite $\mc{H}_{\rm BdG}(\mb{k,q})$ of Eq.\,\eqref{Eq.H_BdG} in the BCS $s$-wave pairing channel in a compact form as, 
\begin{equation}
	\mc{H}_{\rm BdG}^{\rm BCS} (\mb{k}) = \xi(\mb{k}) \pi_z+ \alpha(\mb{k})\sigma_z + J_{sd} \pi_z \sigma_x \tau_z  + \Delta^{\!c} \,\pi_y\sigma_y\ , \label{Eq.BdG_BCS_Ham}
\end{equation}
where, $\xi(\mb{k})=[-2t(\cos k_x + \cos k_y) -\mu]$, $\alpha(\mb{k})=\alpha\sin k_x$, and $\pi$ denotes the Pauli matrices acting on the particle-hole degree of freedom. Note that, $\Delta^{\!c}$ is obtained self-consistently by minimizing the condensation energy density; thus it depends on $J_{sd}$. 

\begin{figure}
	\includegraphics[scale=0.77]{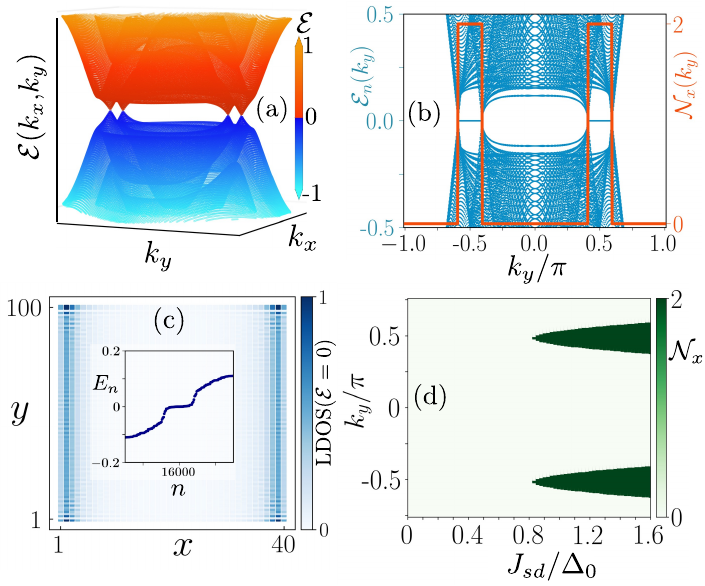} 
	\caption{\tbf{Emergence of topological superconductivity:} \tbf{(a)} Bulk energy spectrum is shown in the TSC phase in the $k_x-k_y$ plane exhibiting four isolated nodal points. \tbf{(b)} Eigenvalue spectrum (left axis) and winding number (right axis) are shown as a function of $k_y$ considering finite size along $x$-direction. \tbf{(c)} LDOS profile at zero energy is shown in the co-ordinate space for the rectangular geometry. Inset displays the variation of energy-eigenvalues $E_n$ as a function of state index, $n$, 
		manifesting the flat zero energy modes. \tbf{(d)} Variation of winding number, $\mc{N}_x$, is shown in the $J_{sd}$-$k_y$ plane. 
		In panels (a)-(c), we choose $(J_{sd},B)=(1.4\Delta_0,0)$ corresponding to the TSC phase of Fig.\,\ref{Fig_2}(a) and set $B=0$ in panel (d).}
	\label{Fig_3}
\end{figure}

In Fig.\,\ref{Fig_3}(a), we depict the bulk energy spectrum $\mc{E}(\mb{k})$ by diagonalizing $\mc{H}_{\rm BdG}^{\rm BCS} (\mb{k})$ for $J_{sd}(=1.4\Delta_0)>J_{sd}^c$. One can observe the 
appearance of four isolated nodal points where the valence and conduction bands touch each other resulting in a gapless superconducting state. The presence of a topological phase is manifested in the bulk-boundary correspondence. To examine this, we next consider a ribbon geometry where the system is periodic along $y$-direction and finite only along $x$-direction with the length $L_x=100a$ ($a$$=1$) being the lattice spacing. The Hamiltonian for this ribbon geometry is obtained by performing an inverse Fourier transformation along $x$-direction (see methods section for details) and the corresponding band structure 
is depicted in Fig.\,\ref{Fig_3}(b) as a function of $k_y$ showcasing the emergence of flat Majorana zero-energy modes connecting the bulk nodal points \ie weak TSC phase.

 To further investigate the localization property of these flat zero-energy modes, we consider the system under the open-boundary condition in both directions i.e., a rectangular geometry being finite size along both $x$- and $y$-direction with $(L_x,L_y)=(60a,100a)$. Then, we again diagonalize the Hamiltonian for this geometry and show the behavior of the eigenvalues $\mc{E}_n$ as a function of state index $n$ in the inset of Fig.\,\ref{Fig_3}(c). Additionally, we compute the site-resolved local density of states (LDOS) at energy $\mc{E}$ using the expression: ${\rm LDOS} (\mb{r},\mc{E})= \sum_n \delta (\mc{E}-\mc{E}_n) |\langle{\mb{r}}|u_{n}(\mb{r})\rangle|^2$, where $\ket{u_n(r)}$ is the eigenvector having eigenvalue $\mc{E}_n$ and $\delta(x)$ is the Dirac-delta function. We illustrate the LDOS (at $\mc{E}=0$) in Fig.\,\ref{Fig_3}(c) as a function of real-space lattice coordinates. From the LDOS profile, it is evident that LDOS is finite and highest in magnitide at the two edges along $y$-direction located at $x=0$ and $x=L_x$, while it vanishes as we proceed towards the bulk of the system. The energy eigenvalues together with the LDOS clearly establish that the Majorana flat zero-energy modes are localized at the boundary of the system.

With these findings of the gapless boundary along with the zero-energy boundary modes, we further investigate the topological origin of these zero-energy flat edge modes by calculating a topological invariant. As mentioned earlier, in the absence of any external Zeeman field, the system retains TRS. Additionally, the charge-conjugation symmetry $\mc{C}=\pi_x\mc{K}$ and  the chiral symmetry $\mc{S}=\pi_x\sigma_y\tau_x$ are also preserved i.e., $\mc{C}^{-1}\mc{H}_{\rm BdG}^{\rm BCS} (\mb{k}) \mc{C} = -\mc{H}_{\rm BdG}^{\rm BCS} (-\mb{k})$ and $\mc{S}^{-1}\mc{H}_{\rm BdG}^{\rm BCS} (\mb{k})\mc{S} = -\mc{H}_{\rm BdG}^{\rm BCS} (\mb{k})$. Thus, our system in the BCS channel belongs to the BDI topological class with a $\mathbb{Z}$ invariant in one-dimension~\cite{ChiuRMP2016}. Topological characterization of the flat edge modes in 2D TSCs can be performed by computing the winding number $\mc{N}_x$ as a function of $k_y$ as ~\cite{ChiuRMP2016,RyuNJP2010,Benalcazar2022_PRL,Pal2025_WSM}:
\begin{eqnarray}
	\mc{N}_x(k_y)=\frac{i}{2\pi} \int_{\rm{BZ}} dk_x \operatorname{Tr}\left[q^{-1}(\mb{k}) \partial_{k_x} q(\mb{k})\right] \label{Eq:winding_num}\ ,
\end{eqnarray}
where, $q(\mb{k})$ is obtained by recasting $\mc{H}_{\rm BdG}^{\rm BCS} (\mb{k})$ into an anti-diagonal form by utilizing the presence of chiral symmetry in the system. Note that, $q(\mb{k})$ is constructed using the self-consistently computed values of $\Delta^{\! c}$ (see methods section). We depict the behavior of  winding number $\mc{N}_x$ as a function of $k_y$ in Fig.\,\ref{Fig_3}(d) and find a perfect one-to-one correspondence between the Majorana flat-edge modes and non-zero $\mc{N}_x$ value, explicitly, $\mc{N}_x=2$. This establishes the emergence of MFEMs as a manifestation of the TSC phase in $p$WMs. 

Before we conclude this discussion, we illustrate the variation of $\mc{N}_x$ in the $J_{sd}- k_y$ plane to identify the parameter regime corresponding to the TSC phase as demonstrated in Fig.\,\ref{Fig_3}(d). It clearly shows the onset of TSC phase at $J_{sd}=0.8\Delta_0$. Our findings are also consistent with the time-reversal symmetric 2D TSC having $(p_x+p_y)$-pairing symmetry and MFEMs proposed in the literature~\cite{Zhang2019_SciRep}. Therefore, we successfully establish the emergence of weak TSC phase in the BCS pairing channel hosting MFEMs without employing any external Zeeman field. Importantly, we do not need any Rashba SOC to obtain this TSC in contrast to the majority of the proposal reported in literature on MFEMs~\cite{PhysRevB.88.180503,Chatterjee2024_PRBLb}.

\begin{figure}
	\includegraphics[scale=0.52]{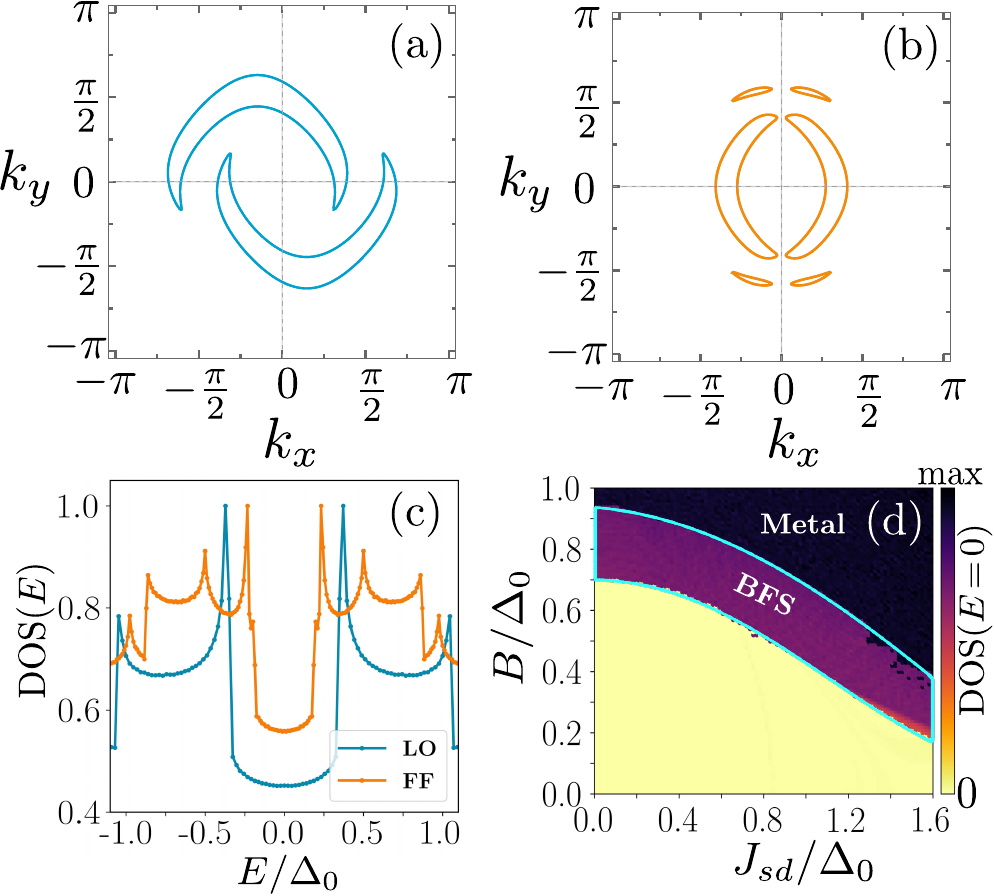} 
	\caption{\tbf{Emergence of BFSs driven by finite-momentum Cooper pairs:} Demonstration of the appearance of BFSs in the $k_x-k_y$ plane corresponding to the gapless \tbf{(a)} LO and \tbf{(b)} FF phase. \tbf{(c)} Normalized single-particle density of states  is shown as a function of $E/\Delta_0$ in the gapless finite-momentum superconductivity channels. \tbf{(d)} Phase diagram of the zero energy density of states is displayed in the $J_{sd}-B$ plane, highlighting the presence of BFSs in the 
		gapless superconducting phases.}
	\label{Fig_4}
\end{figure}

\vskip 0.1cm
\tbf{Bogoliubov Fermi surfaces:}
We now examine the emergence of BFSs as a direct consequence of finite momentum superconductivity.
Since BFSs appear in the gapless superconducting phase, we investigate the presence of BFSs in the gapless LO and also in the gapless FF state. We choose two sets of $(J_{sd}/\Delta_0,B/\Delta_0)$ from the phase diagram of Fig.\,\ref{Fig_2}(a): one from the gapless LO state with $(J_{sd}/\Delta_0,B/\Delta_0)=(0,0.72)$, and the other from the gapless FF state with $(J_{sd}/\Delta_0,B/\Delta_0)=(0.9,0.6)$. We also examine the characteristics of FSs corresponding to these parameter values in Fig.\,\ref{Fig_4}(a) (LO state) and Fig.\,\ref{Fig_4}(b) (FF state), and identify them as BFSs based on the following points. First, the dimensions of these FSs are the same as the underlying normal state FSs, which is one of the defining property of BFSs~\cite{Agterberg2017,Setty2020PRB}. Secondly, there exists a substantial population of Bogoliubov quasiparticles in addition to Cooper pairs and are primarily characterized by the finite density of states around zero energy~\cite{Setty2020PRB}. To establish this, we compute the single particle density of states using ${\rm DOS}(E)=\sum_{\mb{k}} \delta(E-\mc{E}(\mb{k}))$, where $\mc{E}(\mb{k})$ are the energy-eigenvalues of the mean-field BdG Hamiltonian $\mc{H}_{\rm BdG}(\mb{k},\mb{q})$ and show it as a function of $E/\Delta_0$ in Fig.~\ref{Fig_4}(c) for both the LO 
and FF states. To emphasize, the DOS is computed with the self-consistently obtained values of $(\Delta^{\!c},q^c_x,q^c_y)$. Importantly, we find an elevated DOS around the zero energy, indicating the presence of the Bogoliubov quasiparticles as zero-energy excitations. Note that, the elevated DOS is 
larger in magnitude in case of FF state due to the additional satellite structures of BFSs (see Fig.~\ref{Fig_4}(b)). Additionally, we also observe the presence of superconducting coherence peaks at $E/\Delta_0\sim 0.33 $ in the LO state and at $E/\Delta_0\sim 0.2$ for the FF state, which also support the self-consistently obtained values of $\Delta^{\! c}$ in both the LO and FF state. Therefore, our analysis confirms the appearance of BFSs in the gapless LO and FF states driven by the finite-momentum Cooper pairs. 

Now, to demonstrate the appearance of BFSs throughout the gapless LO and FF phases, we compute the DOS $(E=0)$ by scanning the model parameters $(J_{sd},B)$ with the self-consistently obtained values of $(\Delta^{\!c},q^c_x,q^c_y)$ and illustrate in Fig.\,\ref{Fig_4}(d). We observe that the nonzero finite DOS (at $E=0$) appears in the gapless LO, gapless FF, and in the normal state. The nonzero DOS in the normal state appears due to the metallic nature of the underlying normal state system, whereas the finite elevated DOS $(E=0)$ in the LO and FF state represent the emergence of BFSs. Thus, we establish a novel pathway for hosting BFSs driven by the finite-momentum superconductivity .

\tbf{Superconducting diode effect:}
Finally, we focus on another important manifestation of the finite-momentum superconductivity in $p$WMs: the appearance of SDE~\cite{Nadeem2023,LiangFu2022_PNAS}. To achieve this, we compute the supercurrent density, $\mb{J}(\mb{q})$, defined as~\cite{Daido2022_SDE_PRL}, 
\begin{equation}
	J_i(\mb{q}) = 2e\frac{\partial \Omega(q_x,q_y,\Delta_{\mb{q}}^{0})}{\partial q_i}\ ,~i=(x,y) \label{Eq.Jq},
\end{equation} 
where, $e$ is the charge of the electron, $\Omega(q_x,q_y,\Delta_{\mb{q}}^{0})$ is the condensation energy density for a given value of $\mb{q}=(q_x,q_y)$ and $\Delta_{\mb{q}}^{0}$ is the superconducting pairing amplitude obtained self-consistently after minimizing $\Omega(q_x,q_y,\Delta_q^{\rm FF})$ with respect to $\Delta_{\mb{q}}^{\rm FF}$ i.e., $\Omega(\mb{q},\Delta_{\mb{q}}^0)\equiv {\rm min}_{\Delta^{\rm FF}_{\mb{q}}} \Omega(\mb{q},\Delta^{\rm FF}_q)$. Thus, for the actual superconducting ground state, $\mb{q}^c=(q_x^c,q_y^c)$, the supercurrent density $\mb{J}(\mb{q}^c)=0$ by definition. 


To quantify the SDE, we define diode efficiency factor, defined in terms of the critical supercurrent density as,
\begin{equation}
	\eta_i =\frac{|J_i^{c+} - J_i^{c-}|}{J_i^{c+} + J_i^{c-}}\ , ~~ i=x,y
\end{equation} 
where, $J^{c+}_i$ ($J^{c-}_i$) refers to the critical supercurrent densities along the positive (negative) $i$-direction, i.e., $J_i^{c+} \equiv {\rm max}_\mb{q} J_i(\mb{q})$ and $J_i^{c-} \equiv |{\rm min}_\mb{q} J_i(\mb{q})|$. 
For the upcoming discussions on SDE, we follow the convention that the positive (negative) values of $J_{i}(\mb{q})$ represent the flow of supercurrent along the positive (negative) $i$($\in \{x,y\})$-direction. We also emphasize that the SDE is observed only in the FF phase, not in the LO pairing state. This is because in the FF state, the Cooper pairs condense into a unique finite center-of-mass momentum $\mb{q}^c=\{ q_x^c,q_y^c\}$ without having any TRS partner momentum 
at $-\mb{q}^c$. For our system, in the FF state, the condensation energy density, $\Omega(q_x,q_y,\Delta_{\mb{q}}^{0})$ is symmetric with respect to  $q_y$ i.e., $\Omega(q_x,q_y,\Delta_{\mb{q}}^{0})=\Omega(q_x,-q_y,\Delta_{\mb{q}}^{0})$, however asymmetric with respect to $q_x$ i.e., $\Omega(q_x,q_y,\Delta_{\mb{q}}^{0})\ne \Omega(-q_x,q_y,\Delta_{\mb{q}}^{0})$. Consequently, the critical current density along $x$-direction 
becomes nonreciprocal, \ie $J_x^{c+} \ne J_x^{c-}$ leading to nonzero values of only $\eta_x$. In contrast, the critical current density along $y$-direction is completely reciprocal: $J_y^{c+} = J_y^{c-}$. This can be explicitly argued in the following way. If the supercurrent density along $y$-direction attains its maximum value when $\mb{q}=(q_x^m,q_y^m)$ i.e., $J_y(q^m_x,q^m_y)=J_y^{c+}$, then there exists a point with $\mb{q}=(q_x^m,-q_y^m)$ where $J_y(q^m_x,-q^m_y)=2e\frac{\partial \Omega(q_x^m,-q_y^m,\Delta_{\mb{q}}^{0})}{\partial (-q_y^m)}=-2e\frac{\partial \Omega(q_x^m,q_y^m,\Delta_{\mb{q}}^{0})}{\partial (q_y^m)}=-J_y^{c+}$. Thus, the critical supercurrent density along $y$-direction is perfectly reciprocal. Similarly, it can be shown that in the LO pairing state, the behavior of the critical supercurrent density $\mb{J}(\mb{q})$ is reciprocal as $\Omega(q_x,q_y,\Delta_{\mb{q}}^{0})=\Omega(-q_x,q_y,\Delta_{\mb{q}}^{0})=\Omega(q_x,-q_y,\Delta_{\mb{q}}^{0})$. 

\begin{figure}
	\includegraphics[scale=0.52]{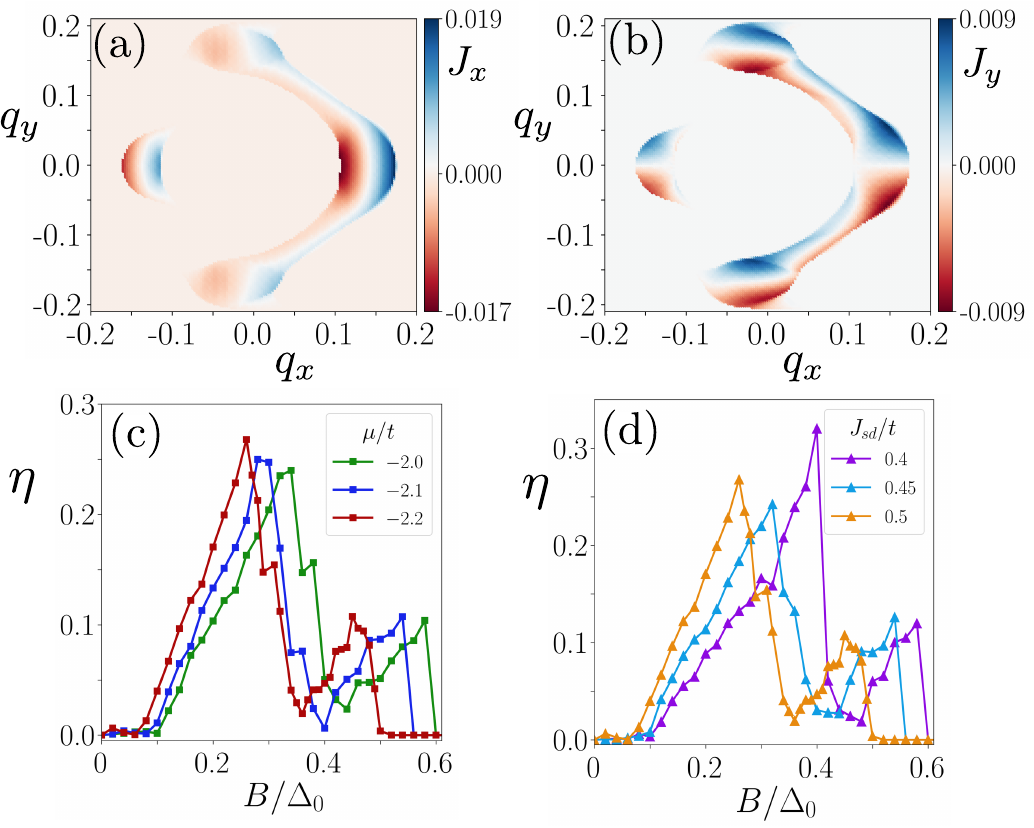} 
	\caption{\tbf{Manifestation of SDE in the FF pairing channel:} In panels \tbf{(a)} and \tbf{(b)}, the variation of supercurrent $J_x(\mb{q})$ and $J_y(\mb{q})$, respectively, are shown in the $q_x-q_y$ plane corresponding to the FF pairing state with $(J_{sd}/\Delta_0,B/\Delta_0)=(0.5,0.7)$. Panels \tbf{(c)} and \tbf{(d)} display the Diode efficiency factor, $\eta$, as a function of the external Zeeman field, $B$ for various choices of $\mu$ and $J_{sd}$, respectively. We choose the other model parameters as $(J_{sd},\alpha)=(0.5t,t)$ in panel (c) and  $(\mu,\alpha)=(-2.2t,t)$ in panel (d).}
	\label{Fig_5}
\end{figure}

We now compute the supercurrent density $\mb{J}(\mb{q})$ numerically using Newton's central difference method and  plot 
and depict the behavior of both $J_x(\mb{q})$ and $J_y(\mb{q})$ in Fig.\,\ref{Fig_5}(a) and (b), respectively, corresponding to the gapless FF phase with $(J_{sd}/\Delta_0,B/\Delta_0)=(0.5,0.7)$. We observe that $J_x^{c+} \ne J_x^{c-}$ but $J_y^{c+} = J_y^{c-}$, as discussed in the previous paragraph, thus, we focus on only $\eta_x$ ($\equiv \eta)$. In Fig.\,\ref{Fig_5}(c), we showcase the variation of $\eta$ as a function of the Zeeman field considering $\mu/t=-2,-2.1,-2.2$ with $J_{sd}=0.5t$. We find the maximum diode efficiency with $\eta=27\%$ for $\mu=-2.2t$. 
We refer to the SM 
for the variation of ($q^c_x,q^c_y,\Delta^{\! c}$) as well as the superconducting bulk gap. 

We further study the variation of $\eta$ for different choices of $J_{sd}$ maintaining fixed $\mu$ and show the behavior in Fig.~\ref{Fig_5}(d). In this parameter space, we find the maximum value of $\eta\sim 33\%$. Note that, in both cases, $\eta$ initially increases with the increase in $B$ upto a critical value of $B$, after which $\eta$ decreases with increasing in $B$. At this critical value of $B$, the system enters into the gapless FF phase from the gapped FF phase (see SM for the details of the bulk BdG gap). This gapless FF phase hosts BFSs as zero-energy excitation, comprised of both the Bogoliubov quasiparticles and Cooper pairs. Thus, density of Cooper pairs is diluted in the gapless state compared to the gapped FF state. This causes the suppression of $\eta$ in the gapless FF state. Interestingly, we also find that within the gapless FF state, $\eta$ increases further with increasing $B$ before vanishes where the superconductivity is completely destroyed. Therefore, from our analysis, we establish the manifestation of finite-momentum superconductivity in the FF state as a mechanism for SDE. Thus, $p$WMs can be
a potential candidate for application in realizing dissipationless superconducting devices. 

                                                                             
{\large{\tbf{Discussion}}}
\vskip -0.08cm
To summarize, we have demonstrated that $p$WMs provide a promising platform for realizing a rich variety of unconventional superconducting phases. These include TSC along with finite-momentum pairing in both FF and LO pairing states. The inherent momentum-dependent spin splitting combined with the intrinsic exchange interaction drives the system into the emergence of topological phases without requiring any external Zeeman field or Rashba spin–orbit coupling, thereby opening routes toward scalable device applications. On the other hand, in the presence of a Zeeman field, the Cooper pairs acquire finite center-of-mass momentum driving the system into a gapless phase hosting BFSs as a zero-energy excitations with potential of applications in quantum devices. Additionally, the emergence of the SDE has also been established in the FF phase, drawing attention as an energy-efficient superconducting device component. On a final note, the superconductivity in our model has been treated within a self-consistent mean-field BdG formalism, ensuring reliable characterization of the superconducting ground state.

In our analysis, we assume an attractive electron-electron interaction in $p$WMs, leading to the emergence of superconductivity. Existence of such phonon mediated interaction is unknown at this moment since very few candidate materials have been proposed \eg~CeNiAsO~\cite{Chakraborty2025}, ${\rm NiI_2}$~\cite{Song2025}, $\rm{Gd_3(Ru_{1-\delta}Rh_\delta)_4Al_{12}}$~\cite{Yamada2025} etc. Nevertheless, there are systems with noncollinear magnetic moments, similar to $p$WMs, where the onset of superconductivity has been experimentally reported such as MnP~\cite{Cheng2015_MnP}, CrAs~\cite{Kotegawa2014} etc. One can also induce superconductivity in $p$WMs by proximitizing to a conventional BCS superconductors such as ${\rm NbSe_2}$. We have analyzed the superconducting pairing correlations originating from the intra-orbital components. Exploring the stability of our proposed phases in presence of inter-orbital pairing terms remains an outlook for our future work.

In conclusion, our work sheds light on the role of newly proposed unconventional $p$WMs in realizing 
the underlying superconducting phases highlighting the intriguing ones with practical applications. 
While our theoretical model simulations were performed in a square lattice, it would be interesting to investigate the same in the Kagome lattice as found for ${\rm CeNiAsCo}$~\cite{Yamada2025}. 
Given the SDE is supported by the finite-momentum superconductivity in static case, one can also study the mechanism and possibility of rectification effect in a superconducting diode via Floquet driving. 


\vskip 0.2cm
{\large{\tbf{Methods}}}
\vskip 0.04cm
{\tbf{Topological invariant}} \vskip 0.08cm
The emergence of TSC phase in the BCS $s$-wave channel has been characterized by the winding number $\mc{N}_x$ ($\mathbb{Z}$ invariant) belonging to the BDI topological class of the ten-fold classification~\cite{ChiuRMP2016}. To compute the winding number, we first recast the BdG Hamiltonian $\mc{H}_{\rm BdG}^{\rm BCS} (\mb{k})$ of Eq.\,\eqref{Eq.BdG_BCS_Ham} into an anti-diagonal form by employing a unitary transformation as,
\begin{eqnarray}
	\tilde{\mc{H}}(\mb{k})= \mc{U}_{s}^{\dagger}\mc{H}_{\rm BdG}^{\rm BCS} (\mb{k}) \mc{U}_{s}=\begin{pmatrix}
		0&q(\mb{k})\\
		q^{\dagger}(\mb{k}) &0\\
	\end{pmatrix} \label{eq:anti_block}
\end{eqnarray}
where the unitary matrix, $\mc{U}_s$, diagonalizes the chiral symmetry operator, $\mc{S}=\pi_x\sigma_y\tau_x$, i.e., $\mc{U}_s^\dagger \mc{S} \mc{U}_s = {\rm diag} (-1,-1,-1,-1,1,1,1,1)$. The $4\times4$ block $q(\mb{k})$ used in Eq.\,\eqref{Eq:winding_num} is given by,
\begin{eqnarray}
	q(\mb{k})=\!\!\begin{pmatrix}
		\xi(\mb{k}) \!-\!\alpha(\mb{k}) & -i\Delta^{\!c} & J_{sd} & 0 \\
		-i\Delta^{\!c} & \xi(\mb{k}) \!-\!\alpha(\mb{k}) & 0 & -J_{sd}  \\
		J_{sd} & 0 & \xi(\mb{k}) \!+\!\alpha(\mb{k}) & -i\Delta^{\!c}   \\
		0 & -J_{sd}& -i\Delta^{\!c}  & \xi(\mb{k}) \!+\!\alpha(\mb{k})   \\
	\end{pmatrix}, \non \\ \label{eq:qk_supp}
\end{eqnarray}
 where, $\xi(\mb{k})$ and $\alpha (\mb{k})$ are as per the definitions given in the main text.
 Using this form of the $q(\mb{k})$ matrix, we numerically compute the winding number, $\mc{N}_x$, taking into account the self-consistently obtained superconducting order parameter.
  
 \vskip 0.2cm
{\tbf{Hamiltonians for finite size geometries.}}\vskip 0.05cm
To illustrate the appearance of MFEMs, we investigate the spectral properties of the BdG Hamiltonian considering both the ribbon and rectangular geometries.
In the ribbon geometry, we choose a finite size system of length $L_x$ along $x$-direction under open boundary conditions, while the system remains periodic along the $y$-direction, thus keeping $k_y$ as a good quantum number. The corresponding Hamiltonian is obtained by performing an inverse Fourier transformation of the Nambu basis along $x$-direction as,
\begin{equation}
	\Psi_{k_x,k_y} = \frac{1}{L_x}\sum_{x=1}^{L_x} e^{ik_xx} \Psi_{x,k_y}\ . \label{Eq.IFT_Ribbon}
\end{equation} 

The mean-field BdG Hamiltonian in the BCS channel (see Eq.~\eqref{Eq.BdG_BCS_Ham}) is given by, 
\begin{eqnarray}
	H_{\rm BdG}\!\!&=&\!\! \sum_{k_x,k_y} \Psi^\dagger_{k_x,k_y} [ \xi(\mb{k}) \pi_z \!+\! \alpha(\mb{k})\sigma_z \!+\! J_{sd} \pi_z \sigma_x \tau_z  \non \\ \!&&\!+ \Delta^{\!c} \,\pi_y\sigma_y] \Psi_{k_x,k_y}\ ,
	\label{Eq. Ham_BCS_Methods}
\end{eqnarray}
We then replace Eq.\,\eqref{Eq.IFT_Ribbon} in Eq.\,\eqref{Eq. Ham_BCS_Methods} and after simplification we obtain the Hamiltonian for the ribbon geometry as, 
\begin{eqnarray}
	H_{\rm BdG}\!\!&=&\!\! \sum_{x=1}^{L_x} \!\sum_{k_y} \Psi^\dagger_{x,k_y} [ (-2t\cos k_y \!-\!\mu)\pi_z \!+\! J_{sd} \pi_z\sigma_x\tau_z \non \\ &+& \Delta^{\!c} \pi_y\sigma_y]\Psi^\dagger_{x,k_y} + \Psi^\dagger_{x,k_y} \left[-t \pi_z + \frac{\alpha}{2i}\sigma_z\right]\Psi^\dagger_{x+1,k_y} \non \\
	&& + {\rm H.c.}\ .   \label{Eq.Ham_Ribbon}
 \end{eqnarray}

In case of the rectangular geometry, we consider a finite size system along both $x$- and $y$-directions with lengths $L_x$ and $L_y$, respectively, under the open boundary condition. Then, we perform the inverse Fourier transformation of the Nambu basis along both $x$- and $y$- directions as, 
\begin{equation}
	\Psi_{k_x,k_y} = \frac{1}{L_x L_y}\sum_{x=1}^{L_x}\sum_{y=1}^{L_y} e^{ik_xx} e^{ik_yy} \Psi_{x,y}\ . \label{Eq.IFT_xy}
\end{equation}
We again replace Eq.~\eqref{Eq.IFT_xy} in Eq.~\eqref{Eq. Ham_BCS_Methods}, and obtain the Hamiltonian for the rectangular geometry as, 
\begin{eqnarray}
	H_{\rm BdG}\!\!&=&\!\! \sum_{x=1}^{L_x} \!\sum_{y=1}^{L_y} \Psi^\dagger_{x,y} [ -\mu\pi_z \!+\! J_{sd} \pi_z\sigma_x\tau_z  + \Delta^{\!c} \pi_y\sigma_y]\Psi^\dagger_{x,y} \non \\ &+& \Psi^\dagger_{x,y} \left[-t \pi_z + \frac{\alpha}{2i}\sigma_z\right]\Psi^\dagger_{x+1,y}  + \Psi^\dagger_{x,y} [-t \pi_z] \Psi^\dagger_{x,y+1} \non \\ 
	&& + {\rm H.c.}\ , \label{Eq.Ham_rectangle}
\end{eqnarray}
We employ the Hamiltonians described in Eqs.\,\eqref{Eq.Ham_Ribbon} and \eqref{Eq.Ham_rectangle} to investigate the emergence of MFEMs and their edge localization in the TSC phase.

\vskip 0.2cm
{\large{\tbf{Data Availability}}}
\vskip 0.08cm
The datasets generated and analyzed during the current study are available from the authors upon reasonable request.

\vskip 0.2cm
{\large{\tbf{Code Availability}}}

The codes and data generated during this study are available from the authors upon request.

\vskip 0.2cm
{\large{\tbf{Acknowledgements}}}
\vskip 0.08cm

A.P. acknowledges Ganpathy Murthy, Mathias S. Scheurer, Arnob Kumar Ghosh, Suman Jyoti De, and Moonsun Pervez for stimulating discussions. A.P. and A.S. acknowledge the SAMKHYA: High-Performance Computing facility provided by the Institute of Physics, Bhubaneswar and the two Workstations provided by the Institute of Physics, Bhubaneswar from the DAE APEX Project for numerical computations. P.D. acknowledges the Department of Space (DoS), India for all support at Physical Research Laboratory (PRL).

\vskip 0.2cm
{\large{\tbf{Authors contributions}}}
\vskip 0.08cm
P.D. and A.S. jointly conceived the idea and supervised the project. A.P. performed all calculations and prepared the manuscript with input from P.D. and A.S. All authors contributed to the science discussions and manuscript preparation.

\bibliographystyle{apsrev4-2mod}
\bibliography{bibfile.bib}
%
\clearpage	
\normalsize\clearpage
\begin{onecolumngrid}
	\begin{center}
		{\fontsize{10.0}{10.5}\selectfont
			\textbf{Supplementary Material for ``Emergent superconducting phases in unconventional $p$-wave magnets: Topological superconductivity, Bogoliubov Fermi surfaces and superconducting diode effect}''\\
			{\normalsize  Amartya Pal\orcidA,$^{1,2}$ Paramita Dutta\orcidB{},$^{3}$ and  Arijit Saha\orcidC{},$^{1,2}$ \\[1mm]}
			{\small $^1$\textit{Institute of Physics, Sachivalaya Marg, Bhubaneswar-751005, India}\\[0.5mm]}
			{\small $^2$\textit{Homi Bhabha National Institute, Training School Complex, Anushakti Nagar, Mumbai 400094, India}\\[0.5mm]}
			{\small $^3$\textit{Theoretical Physics Division, Physical Research Laboratory, Navrangpura, Ahmedabad-380009, India}\\[0.5mm]}
		}
	\end{center}
	
		\newcounter{defcounter}
\setcounter{defcounter}{0}
\setcounter{equation}{0}
\renewcommand{\theequation}{S\arabic{equation}}
\setcounter{page}{1}
\pagenumbering{roman}

\renewcommand{\thesection}{S\arabic{section}}

\tableofcontents 
	
	\section{Transition from gapped to gapless superconducting phase driven by finite-momentum Cooper pairs} 
	\label{sec:Gap_to_gapless}
	In the main text, we have proposed that the finite-momentum Cooper pairs drive the system from a gapped to a gapless superconducting phase with both Fulde-Ferrel (FF) and Larkin–Ovchinnikov (LO) pairings. These gapless FF and LO pairing states host the unique Bogoliubov Fermi surfaces (BFSs) as zero-energy excitations. 
	In this section, we further provide some additional results to support our findings of the main text. 
	
	We illustrate the behavior of the bulk superconducting gap with the variation of the external Zeeman field in Fig.~\ref{Fig_S1}. Notably, we consider the mean-field order parameter obtained self-consistently by minimizing the condensation energy density. The mean-field Bogoliubov-de Gennes (BdG) Hamiltonian (Eq.~(9) of the main text) in the finite momentum pairing channel is given by, 
	\begin{equation}
		\mc{H}_{\rm BdG}(\mb{k},\mb{q^c}) = \begin{bmatrix}  
			\mc{H}_{\rm N}(\mb{k}+\frac{\mb{q^c}}{2}) & -i\sigma_y \Delta^{\!c}\\  
			i\sigma_y \Delta^{\!c} & -\mc{H}_{\rm N}^T(-\mb{k}+\frac{\mb{q^c}}{2}) 
		\end{bmatrix}\ ,		\label{Eq.H_BdG}
	\end{equation}
	\vspace{0.1cm}
	\begin{figure}[h]
		\includegraphics[scale=0.6]{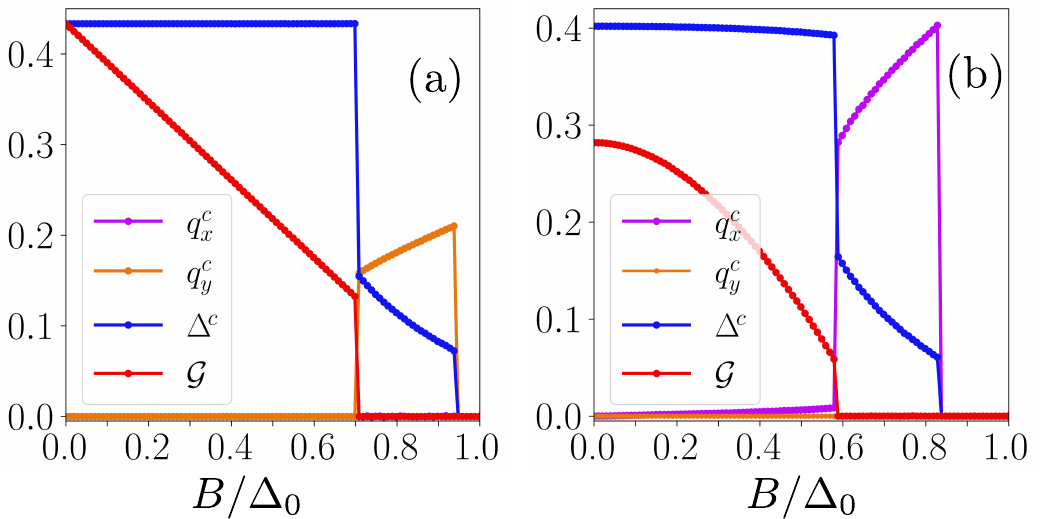} 
		\caption{Variation of superconducting order paramters, $(q_x^c,q_y^c,\Delta^{\!c})$ and bulk BdG gap, $\mc{G}$ are shown as a function of the external Zeeman field, $B/\Delta_0$ in the LO (panel \tbf{(a)}) and FF pairing channels (panel \tbf{(b)}). We set $J_{sd}/\Delta_0=0$ (panel \tbf{(a)}) and $J_{sd}/\Delta_0=0.6$ (panel \tbf{(b)}). Other model parameters are chosen as $(\alpha,\mu,t)=(t,-2t,t)$.}
		\label{Fig_S1}
	\end{figure}
	which corresponds to the ground state characterized by $(q_x^c,q_y^c,\Delta^{\! c})$ and thus it depends on the parameters of the Hamiltonian. 
	
	We first diagonalize the above Hamiltonian to obtain the energy eigenvalues, $\mc{E}(\mb{k})$, for each momentum point $\mb{k}=\{k_x,k_y\}$. Then, we compute the bulk gap as $\mc{G} = 2\,{\rm min}_{\mb{k}}|\mc{E}(\mb{k})|$. We then depict the variation of $\mc{G}$ as well as $(q_x^c,q_y^c,\Delta^{\! c})$ as a function of the Zeeman field $B/\Delta_0$ in the LO pairing channel (see Fig.\,\ref{Fig_S1}(a)) and FF pairing channel (see Fig.\,\ref{Fig_S1}(b)). In the LO phase, 
	we observe that the bulk gap of the BdG eigenvalue spectrum becomes gapless when $q^c_y\ne0$ but $q^c_x=0$. Remarkably, both $\Delta^{\!c}$ and $q_y^c$ exhibits discontinuous behaviours when the system enters into the gapless phase. The value of the $\Delta^{\!c}$ is also reduced in this gapless phase. 
	On the other hand, in the FF phase, the system initially remains gapped even the Cooper pairs acquire a small but finite momentum with $q^c_x\ne0$ but $q^c_y=0$. Then, the system becomes gapless associated with a discontinuous change in $\Delta^{c}$ and $q_x^c$ only after a certain strength of $B/\Delta_0$. The amplitude of $q_x^c$ also grows further in the gapless phase. This behavior is consistent with the behavior observed in the LO and FF pairing channels supporting our discussions in the main text.
	
	\section{Behavior of superconducting order parameter}~\label{sec:sc_order}
	In the main text, we have illustrated the behavior of the superconducting diode efficiency $\eta$ in the FF pairing channel by varying the external Zeeman field for various choices of $\mu$ and $J_{sd}$ (see Figs.\,5(c-d) in the main text). For the sake of completeness and understanding, here we depict the variation of the mean-field superconducting order parameter $\Delta^{\!c}$ and also $q^c_x$ for the same set of model parameters. 
	
	In Figs.~\ref{Fig_S2}(a) and (c), we depict the variation $\Delta^{\!c}$ and $q^c_x$, respectively as a function of the external Zeeman field, for three choices of chemical potential $\mu$ and $J_{sd} \ne 0$, $\alpha \ne 0$, while, the variation of $\Delta^{\!c}$ and $q^c_x$ are shown for three choices of the exchange strength $J_{sd}$ considering $\mu\ne 0$ and $\alpha \ne 0$ in Figs.~\ref{Fig_S2}(b) and (d) respectively. The behavior of $\Delta^c$ remains consistent and qualitatively similar for different values of the chemical potential before it becomes zero after the critical magnetic field as presented in Fig.~\ref{Fig_S2}(a). These phenomena remain qualitatively similar irrespective of the exact values of exchange couping as shown in Fig.~\ref{Fig_S2}(b). 
	From Figs.~\ref{Fig_S2}(c) and (d), we observe that the Cooper pair momentum, $q^c_x$, initially increases linearly with the Zeeman field but it changes discontinuously when the system enters into the gapless phase supporting our dicsussions in Sec.~\ref{sec:Gap_to_gapless}. As a consequence, the diode efficiency decreases significantly in the gapless FF phase compared to the gapped FF superconducting phase as discussed in the main text (see Figs.\,5(c-d)). We also note that the maximum diode efficiency appears close to the point where the bulk becomes gapless. These behaviors are fully consistent with our analysis described 
	in the main text.
	\vspace{0.2cm}
	\begin{figure}[h]
		\includegraphics[scale=0.6]{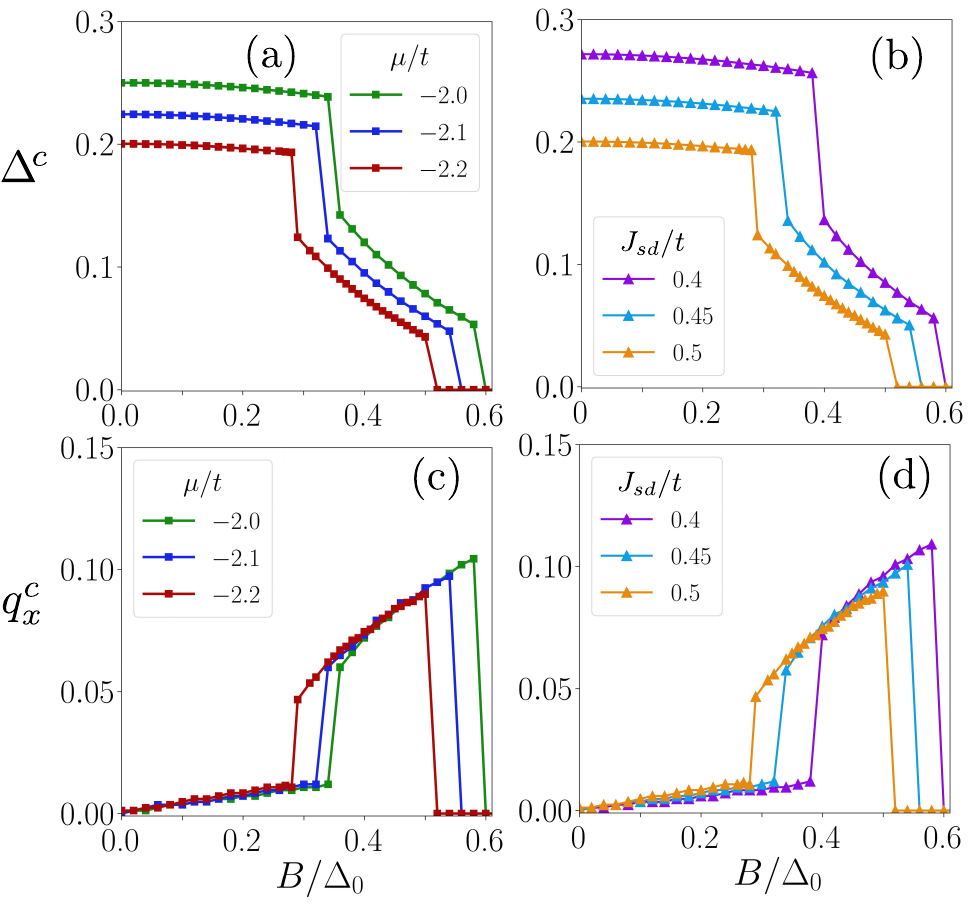} 
		\caption{Variation of  $\Delta^{\!c}$ [(a)-(b)] and  $q_x^c$ [(c)-(d)] are displayed as functions of Zeeman field $B/\Delta_0$ for various choices of the model parameters. In panels (a), (c), we set $(J_{sd},\alpha)=(0.5t,t)$ whereas in panels (b), (d) we choose $(\mu,\alpha)=(-2.2t,t)$.}
		\label{Fig_S2}
	\end{figure}
	
\end{onecolumngrid}	

\end{document}